\documentclass{elsart}
\usepackage{amsmath,amsfonts,amssymb,natbib}
\usepackage{graphicx} 
\usepackage{color} 
\journal{Physica A}
\def\lm {\lambda}
\begin{document}
\begin{frontmatter}
\title{Reversible Adsorption on a Random Site Surface}
\author{C. Oleyar and J. Talbot}

\address{Department of  Chemistry  and  Biochemistry,
Duquesne University, Pittsburgh, PA 15282-1530, USA}
\bibliographystyle{elsart-num}

\begin{abstract}
  We examine the reversible adsorption of hard spheres on a random
  site surface in which the adsorption sites are uniformly and
  randomly distributed on a plane. Each site can be occupied by one
  solute provided that the nearest occupied site is at least one
  diameter away. We use a numerical method to obtain the adsorption
  isotherm, i.e. the number of adsorbed particles as a function of the
  bulk activity. The maximum coverage is obtained in the limit of
  infinite activity and is known exactly in the limits of low and high
  site density. An approximate theory for the adsorption isotherms,
  valid at low site density, is developed by using a cluster expansion
  of the grand canonical partition function. This requires as input
  the number of clusters of adsorption site of a given size. The
  theory is accurate for the entire range of activity as long as the
  site density is less than about 0.3 sites per particle area. We also
  discuss a connection between this model and the vertex cover
  problem.
\end{abstract}


\end{frontmatter}

\section{Introduction}

The adsorption of proteins and colloids at a liquid-solid interface is
a key step in many natural and industrial processes such as
filtration, chromatography, protein purification, immunological
assays, biosensors, biomineralization and biofouling. In many of these
situations, the surface of the adsorbent is heterogeneous. For
example, in immunological assays one typically employs colloidal
particles that have been coated with proteins (antibodies) to bind
with antigens that may be present in the sample.  Similarly, in affinity
chromatography, the adsorbent is synthesized by immobilizing certain
affinity ligands on porous silica, agarose or synthetic polymers. Many
of these applications can benefit from a quantitative knowledge of the
amount of solute that is adsorbed as a function of the bulk
concentration, ligand density and distribution and solute properties
such as the size.

The modeling approach required depends on the nature of the
adsorption. When there is a finite desorption probability,
characterized by a non-zero desorption rate constant, an equilibrium between the
bulk and adsorbed phases will be established; rapidly if the
desorption rate is large and more slowly for small desorption
rates. The properties of the equilibrium state, in particular the
adsorption isotherm, depend on the bulk phase activity. If the
desorption rate constant is very small on the experimental time scale
the adsorption is effectively irreversible and a different approach is
required.

The statistical mechanics of reversible adsorption on
heterogeneous surfaces, particularly for gases
on solids, has a long history. Pioneering work was
performed by Hill \cite{H49} and
Steele \cite{S63} and it is still an active area of research
\cite{RE92,S99}.  There is also a
well-developed literature on irreversible adsorption on homogeneous surfaces
\cite{E93,TTVV00,SVS00}. Some models specifically address irreversible
adsorption on non-uniform surfaces \cite{JWTT93,JTW94,AWM02,ASWM02}  
where
macromolecules are represented as hard spheres that bind irreversibly to adsorption sites. 
In the simplest of these, the Random Site Surface (RSS) \cite{JWTT93}, 
the sites are represented by randomly distributed
points. Adamczyk et al \cite{AWM02} extended the basic model
to the situation where the adsorption sites have finite dimensions.
 The adsorption of colloidal 
\cite{LRB98,AJSW04} and nanoparticles \cite{KOAD04} has been
interpreted with these models and a similar
hard sphere model was used to
rationalize the adsorption of proteins to hydrophobic sites on mixed
self-assembled monolayers \cite{OGMRW03}.

Although macromolecules such as colloids and proteins have a tendency
to adsorb irreversibly, this is not always the case and it is
certainly useful to understand the equilibrium behavior. It this
article we therefore present numerical and approximate analytical
results for reversible adsorption of hard spheres on the RSS. For
irreversible adsorption on this surface the task of
developing a theoretical description was greatly simplified by the
existence of an exact mapping to an analogous process on a continuous
(homogeneous) surface \cite{JWTT93}. For reversible adsorption on the
RSS surface, however, there appears to be no similar mapping that
would allow us to exploit the known behavior of hard spheres on
continuous surfaces \cite{T36,T97}.

In addition to its application to adsorption, the RSS model is also
interesting from another perspective. Weight and Hartmann obtained
analytical results for the minimal vertex cover on a random graph by a
mapping to a hard sphere lattice gas \cite{WH00,WH01}. A vertex cover
of an undirected graph is a subset of the vertices of the graph which
contains at least one of the two endpoints of each edge. In the vertex
cover problem one seeks the {\it minimal vertex cover} or the vertex
cover of minimum size of the graph. This is an NP-complete problem
meaning that it is unlikely that there is an efficient algorithm to
solve it. The connection to the adsorption model is made by
associating a vertex with each adsorption site. An edge is present
between any two vertices (or sites) if they are closer than the
adsorbing particle diameter. The minimal vertex cover corresponds to
densest particle packings. Weight and Hartmann obtained an analytical
solution for the densest packing of hard spheres on random graph using
a replica-symmetric approach. Although the random graph is related to
the RSS model, it is not the same as the former does not
have a physical structure. Specifically, in a random graph each
possible edge is present with a given probability, $c$. In the random
site surface, on the other hand, two adsorption sites that are
neighbors of a given site are more likely to be neighbors of each
other than two randomly selected sites. This effect can be quantified
by the clustering coefficient, which is the average probability that
two neighbors of a given vertex are also neighbors of one another.

In section 2 we define the model and the simulation procedure. Section
3 compares numerical results for the maximum coverage with various
theoretical estimates. The structure of the random site surface is
discussed in section 4, and section 5 presents an approximate theory
for the adsorption isotherms that applies at low adsorption site
density.

\section{Model and Simulation}

\begin{figure}
\begin{center}

\resizebox{10cm}{!}{\includegraphics{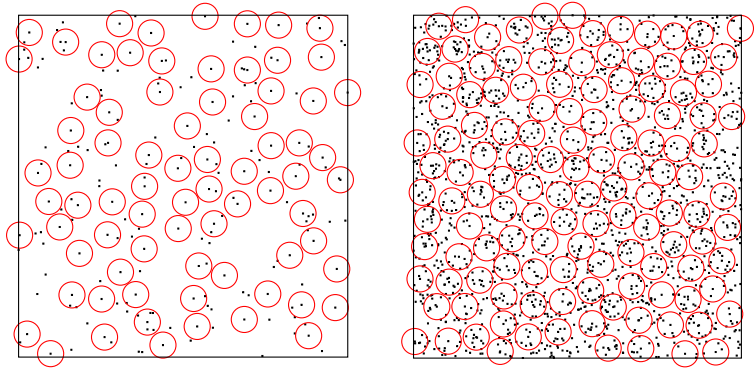}}
\caption{Sample configurations of adsorbed hard spheres on the random site surface. A sphere may
bind, centered, to a site as long as the nearest occupied site is at least $\sigma$ away.  Left configuration: $\alpha=1, N_s=200, \theta=0.380$; right configuration: $\alpha=10,N_s=2000, \theta=0.650$. In both cases $\lm=10000$. }\label{fig:config}
\end{center}
\end{figure}

The adsorbent surface consists of $N_s$ point sites that have been
uniformly and randomly distributed on a surface of area $L^2$. These
sites are frozen in place.  The adsorbate molecules are represented by
hard spheres of diameter $\sigma$. In order to adsorb, a hard
sphere must bind, centered, to one site. An adsorbed sphere
may cover any number of sites but only occupies (or interacts with)
the one at its center. Any site that lies within a distance $\sigma$ of
an occupied site is unavailable. See Figure 1.  It is
convenient to introduce a dimensionless site density
\begin{equation}
\alpha=\frac{\pi\sigma^2N_s}{4L^2}
\label{eq:alpha} 
\end{equation}
corresponding to the average number of sites
in an area equal to the projected area of a sphere.
The  coverage is defined as
\begin{equation}
\theta=\frac{\pi\sigma^2n}{4L^2},
\label{eq:theta}
\end{equation}

where $n$  is the number of adsorbed spheres. 
If the adsorption is irreversible it was shown in Ref
\cite{JWTT93} that the coverage at time $t$ is equal to the coverage on
a continuous surface at time $\tau$,

\begin{equation}
\theta(t;\alpha)=\theta_c(\tau),
\end{equation}
where $\tau$ and $t$ are related by
\begin{equation}
\tau=\alpha(1-e^{-t/\alpha}),
\end{equation}
which has the required property that as $\alpha\to\infty$, $\tau\to t$. 
Although the time-dependent coverage on
the continuous surface $\theta_c(\tau)$, is itself not known exactly, semi-empirical
equations that provide an accurate representation of numerical
simulations of the model have been proposed \cite{JWTT93}.  It is not clear, however,
how to develop a similar mapping to apply to the reversible (equilibrium) case.

In   the reversible binding case there is an equilibrium between a bulk
phase containing adsorbate at activity $\lambda$, and the 
thermodynamic properties of the adsorbed phase can
be obtained from the grand canonical partition function. For a particular
realization of the adsorption site distribution this is:

\begin{equation}
\Xi(\lambda,N_s)=\sum_{n_i=0,1}[\prod_{i>j}(1-f_{ij}n_in_j)\prod_i \lambda^{n_i}],
\end{equation}

where $\lambda=\exp(\beta\mu)$ is the activity with $\beta=1/k_BT$ and $\mu$ is the
chemical potential. Site $i$ is occupied (vacant) if $n_i=1(0)$ and $f_{ij}=1$ if
sites $i$ and $j$ are closer than $\sigma$ and zero otherwise.

The number of adsorbed molecules can be computed directly from the
partition function:
\begin{equation}
<n>=\lambda\left(\frac{\partial\ln \Xi}{\partial\lambda}\right)_{N_s}
\label{eq:nbar1}
\end{equation}
as can the fluctuation in the adsorbed number:
\begin{equation}
<(\delta n)^2>=\lambda\left(\frac{\partial<n>}{\partial\lambda}\right)_{N_s}
\end{equation}

It is generally not possible to evaluate this partition function
analytically, even in one-dimension. However, if all the sites are
isolated $f_{ij}=0, \forall i,j$ the partition function becomes
\begin{equation}
\Xi(\lambda,N_s)=(1+\lambda)^{N_s},
\end{equation}
which yields the Langmuir form for the fraction of occupied sites, 
\begin{equation}
\theta=\alpha\frac{\lambda}{1+\lambda}
\end{equation}
and fluctuation
\begin{equation}
<(\delta n)^2>=N_s\frac{\lambda}{(1+\lambda)^2}.
\end{equation}

We have used a Gillespie type algorithm \cite{BKL75,G77} to obtain the
adsorption isotherms of the RSS model numerically. First, $N_S$ sites are
distributed uniformly and randomly in a periodic unit cell. A list of
the neighbors, i.e. those that lie within a distance $\sigma$, of each site
is then constructed. At each step of the simulation, the total event
rate is $R=\phi+n/ \lm$ where $\phi$ is the fraction of sites that are
available, and $n$ is the number of occupied sites. A waiting time is
generated from an exponential distribution using $t=-\ln(\xi)/R$ where
$\xi$ is a uniform random number. A second random number is used to
decide the type of event. If $\xi<\phi/(\phi+n/ \lm)$ a new sphere is added to
the surface at an available site selected at random. Otherwise, one of
the spheres on the surface is removed. For a given value of $\lm$, the
adsorption/desorption process is allowed to continue until the system
reaches a steady state and then the number of adsorbed spheres is
averaged over a large number of events (typically one million). In the
steady state this procedure is equivalent to a grand canonical Monte
Carlo simulation.

All of the simulation results reported here were obtained using 4000
adsorption sites. Several tests were performed with larger and smaller
systems, but there was no noticeable system size effect. Up to 20
different realizations of the adsorption sites were used.  The
isotherms were generated by performing a series of simulations with
increasing activity in the interval $\lm_{\rm min}\leq\lm\leq\lm_{\rm
  \max}$.  The final configuration of a simulation at given activity
served as the initial configuration for the following simulation at
higher activity. For each value of the activity $n_{\rm trial}$ events
were simulated. In order to allow the system to equilibrate at each
new activity, only the second half of the ntrial events were used in
the calculation of the ensemble averages.

The activity schedule was determined from the formula:

\begin{equation}
\lm_i = \lm_{\rm min}\bigl(\frac{\lm_{\rm max}}{\lm_{\rm min}}\bigr)^{\frac{i}{n_{\lm}-1}},\;\;i=0,\ldots,n_{\lm}-1
\end{equation}

For the adsorption isotherms at low activities we used $\lm_{\rm min}=0.1, \lm_{\rm max}=100$ and $n_{\lm}=20$ and $n_{\rm trial}=50000$. 

As long as the activity is not too large, or whenever $\alpha\leq 0.8$, the
simulations are non-problematic in that the steady state is reached
rapidly and reproducibly. For larger values of $alpha$ at high
actvities, some care is needed to ensure that the system reaches
equilibrium. The parameters $n_\lm$ and $n_{\rm trial}$ were incresed
until further changes had no influence of the isotherms. In most
cases, $n_\lm=50$ and $n_{\rm trial}=5x10^{5}$ were sufficient.

Although we did not employ the method in this study,
parallel tempering \cite{G91,ED05} would be an alternative, and possibly more
efficient, method at large values of the activity.

\begin{figure}
\begin{center}

\resizebox{8cm}{!}{\includegraphics{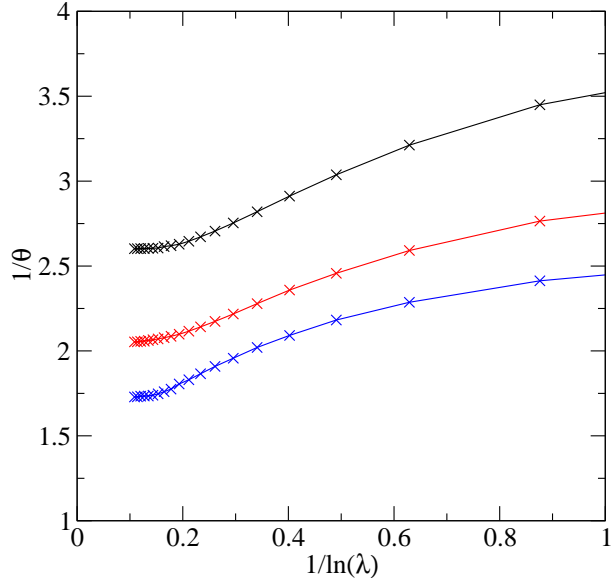}}
\caption{Adsorption Isotherms from simulation. Results are averages over 20 realizations of 4000 sites. $\alpha=1,2,4$ top to bottom.}\label{fig:isotherms}
\end{center}
\end{figure}

\section{Maximum Coverage}

The maximum coverage is obtained in the limit $\lm\to \infty$.
As as the site density approaches zero, all sites become independent
and the maximum coverage is $\alpha$. In the limit $\alpha\to \infty$
the surface becomes smooth and an adsorbing sphere can occupy any
position. The densest configuration  in this case corresponds to a 
hexagonal close packing of spheres with a coverage of
$\theta_{\rm hcp}=\pi/(2\sqrt{3})=0.90687...$ \cite{f64}.

\begin{figure}
\begin{center}

\resizebox{8cm}{!}{\includegraphics{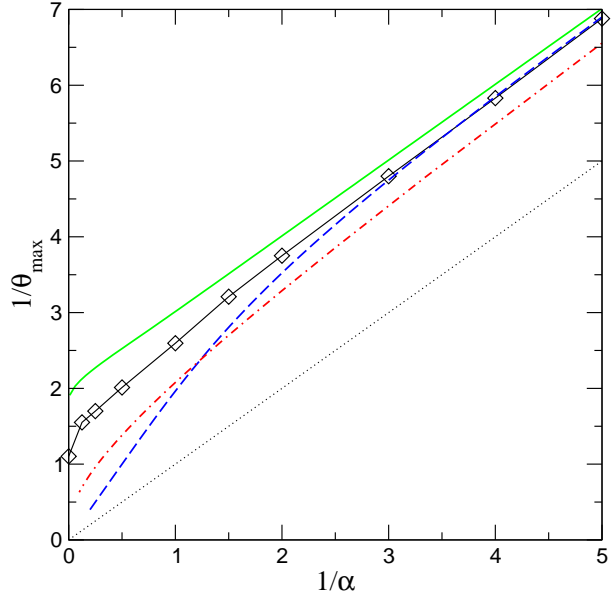}}
\caption{Maximum Coverage as a function of the inverse site density. Squares: numerical simulation; upper solid line: irreversible adsorption; dashed line: cluster theory to second order; dash-dotted line: 
random graph; dotted line: independent  site assumption.}\label{fig:maxcov}
\end{center}
\end{figure}

To obtain numerical values between these two limits for a given value
of $\alpha$ we performed simulations using $\lm_{\rm min}=10, \lm_{\rm max}=100000$ and $n_{\lm}=50$. Some sample isotherms, showing $1/ \theta$ versus $1/ \lm$,
are shown in Figure \ref{fig:isotherms}.  When $1/ \lm$ is
sufficiently small, the curve approaches a plateau. These maximum
values for different $\alpha$ are presented in Fig
\ref{fig:maxcov} together with various theories including the Langmuir
approximation,  the
cluster theory to second order (see section 5) and the maximum coverage in an
irreversible adsorption process (the ``jamming limit'') \cite{JWTT93}.

Weight and Hartmann \cite{WH00,WH01} used a statistical mechanics
approach to the minimum vertex cover problem on a finite connectivity
random graph. By using a replica-symmetric approach, they showed that
the average maximum fraction of occupied graph vertices is
\begin{equation}
\nu_{\rm max}=\frac{2W(c)+W(c)^2}{2c},
\label{eq:whc}
\end{equation}
where $W(x)$ is the Lambert-W function defined by $W(x)e^{W(x)}=x$ and
$c$ is the average connectivity of the random graph. From Eq
\ref{eq:alpha} and Eq \ref{eq:theta} it is easy to show that
$\nu=N/N_s=\theta/ \alpha$. Finally, in order to connect $c$ to $\alpha$ we note
that in a random graph, the distribution of connectivities is given by
a Poisson distribution with mean $c$. Therefore, the probability that
a vertex has a connectivity of zero is $\exp(-c)$. Similarly, for a
random distribution of points in the plane at a density $\alpha$ the
probability that a point has no neighbors (closer than $\sigma$) is
$\exp(-4\alpha)$ (c.f. Eq \ref{eq:x1}). Therefore, we take $c=4\alpha$.
Substituting these results in Eq \ref{eq:whc}, the maximum coverage on
a random graph is
\begin{equation}
\theta_{\rm max}=\frac{1}{4}W(4\alpha)+\frac{1}{8}W(4\alpha)^2.
\label{eq:tmrg}
\end{equation}

It is evident
that the Langmuir approximation of independent sites is very poor, even
for quite small values of $\alpha$. As expected, due to the inherent
inefficiency of irreversible adsorption, the RSA coverage
underestimates the maximum coverage for all values of $\alpha$. The
remaining theories, all of which are equilibrium, consistently
overestimate the maximum coverage. The cluster theory to second order
provides a good estimate up to $\alpha\approx0.3$

\section{Structure of the random site surface}

\begin{figure}
\begin{center}

\resizebox{11cm}{!}{\includegraphics{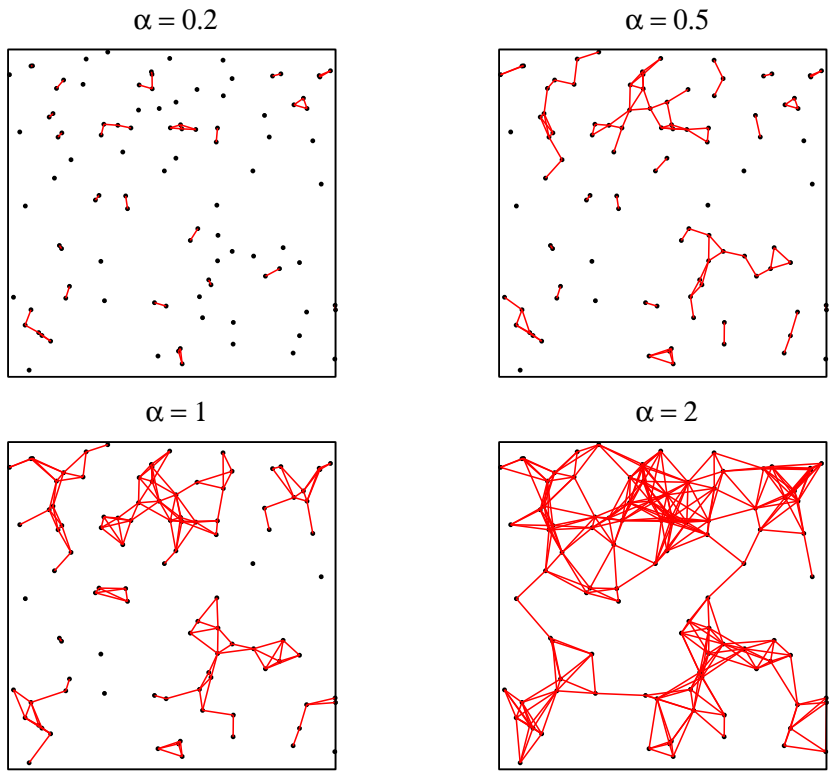}}
\caption{Structure of the random site surface for different values of $\alpha$. Two sites are connected if they are closer than $\sigma$.}\label{fig:structure}
\end{center}
\end{figure}

The connectedness of the adsorption sites is a strong function of the
dimensionless site density, as can be seen in Fig.
\ref{fig:structure}.  For a given distribution of sites, increasing
this parameter corresponds to increasing the size of the adsorbing
spheres. At small values of $\alpha$, the surface consists of isolated
clusters of sites. A cluster consists of $n$ sites all of which are
closer than $\sigma$ to at least one other site of the cluster.  For small
values of $\alpha$ the majority of sites are isolated.  Above a certain
value of $\alpha_c\approx0.8$ percolation occurs 
and almost all sites belong to a giant cluster \cite{G61}.

\begin{figure}
\begin{center}

\resizebox{7.5cm}{!}{\includegraphics{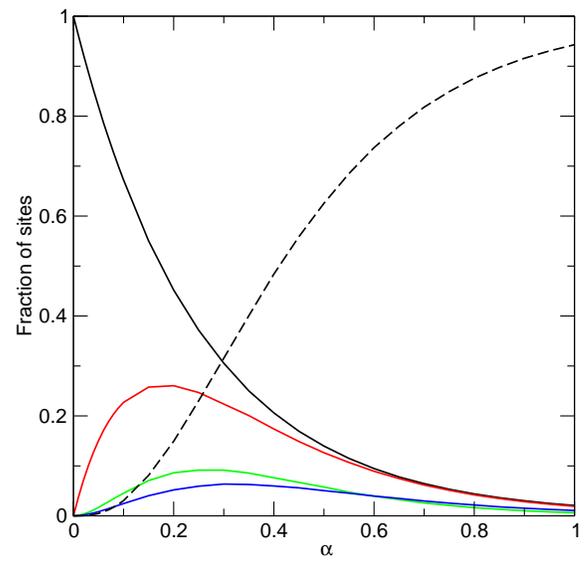}}
\caption{Fraction of sites belonging to clusters of various sizes. 
From top to bottom, left hand  side: isolated  sites, pairs, triplets type a; triplets type b. 
The dashed line shows the fraction of sites not belonging to any of these types}\label{fig:clusterdist}
\end{center}
\end{figure}

In  order to develop an approximate theory for the adsorption isotherms,
we need to know the expected number of clusters of given type as
a function of the site density. Analytical expressions for isolated sites
and pairs can be obtained by making use of the following well-known result. 
Consider a region of area $A$ that
contains a finite number $N$ of sites at given positions.  The probability
that none of the remaining $N_s-N$ points lies in this region (in the large 
$N_s$ limit) is
\begin{equation}
p_0=e^{-\rho_sA}.
\end{equation}
where $\rho_s=N_s/L^2$ is the site density. 
A given site is isolated if there are no other sites within a circle of radius $\sigma$
centered on the given site. 
Therefore the expected number of isolated sites (per adsorption site) is
\begin{equation}
x_1=e^{-4\alpha}.
\label{eq:x1}
\end{equation}
The number of pairs of sites that are separated by a distance between
$r$ and $r+dr$ is
$\frac{1}{2}2\pi r\rho_sdr$. Now imagine two circles of radius $\sigma$ centered on these sites. 
Let $A_2(r)$ represent the area of union of these two circles
If any other site lies  within this area, the  two sites
do not form a pair. The probability that no other site lies within this 
area is $\exp(-A_2(r)\rho_s)$ where  of
diameter $\sigma$ separated by $r$. So the probability that two sites separated by $r$ 
form a pair is $\rho_s \pi r \exp(-A_2(r)\rho_s)dr$. Integrating over $r$ and introducing 
the dimensionless variables $\alpha$ and $y=r/ \sigma$ leads to the final expression for the number of pairs:
\begin{equation}
x_2=4\alpha\int_0^{1}\exp\bigl[-\frac{4\alpha}{\pi}A_2(y)\bigr]ydy,
\label{eq:npair} 
\end{equation}
where $A_2(y)=2(\pi -\cos^{-1}(y/2))-y\sqrt{1-y^2/4}$.  Although this
integral cannot be expressed in terms of elementary functions, it is
straightforward to calculate numerically. A similar procedure could
obviously be used to evaluate the expected number of higher order
clusters, although the task rapidly becomes cumbersome.

Numerical results for isolated sites, and clusters composed of two and
three sites are presented in Fig \ref{fig:clusterdist}. There were
obtained by averaging over 100 independent distributions of 4000 sites
each in a square cell with periodic boundary conditions.  The theoretical expression, Eq
\ref{eq:npair} is indistinguishable from the numerical results. 

For $\alpha\leq0.1$ the surface consists almost entirely of
isolated sites and pairs, while for $\alpha=0.5$ about 50\% of the sites
belong to clusters of cardinality four or greater. For convenience, we
also fitted the numerical results using simple functions. For pairs
the numerical data is very well fitted by the equation:
\begin{equation}
x_2=2\alpha \exp(-5.578\alpha),\; 0\leq \alpha \leq 1.
\label{eq:fit2}
\end{equation}
There are two kinds of triplets. In type ``3a'' two sites are neighbors of a third
site but not of each other, while in type ``3b'' each site is a neighbor of the other
two. The number of each kind is well described by
\begin{equation}
x_{3a}=3.249\alpha^2\exp(-7.537\alpha),\; 0\leq \alpha \leq 1
\label{eq:fit3a}
\end{equation}
and
\begin{equation}
x_{3b}=1.519\alpha^2\exp(-6.250\alpha),\; 0\leq \alpha \leq 1.
\label{eq:fit3b}
\end{equation}

\begin{figure}[t]
\begin{center}

\resizebox{8cm}{!}{\includegraphics{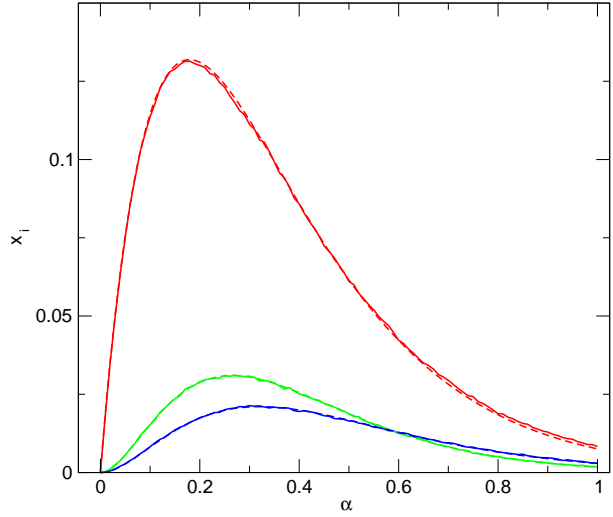}}
\caption{Fraction of pairs, and triplets of type 3a and 3b (top to bottom left hand side) as a function of the dimensionless site density from numerical simulation. 
The fitting functions Eqs \ref{eq:fit2},\ref{eq:fit3a}, and \ref{eq:fit3b} are shown as dashed lines.}\label{fig:cluster3}
\end{center}
\end{figure}

As can be see  in Fig
\ref{fig:cluster3}, these fitting functions provide an excellent approximation of the numerical results. 

\section{Approximate theory for the adsorption isotherms at low site density}

Assuming that the adsorption surface consists of isolated clusters of
sites, the partition function may be expressed as
\begin{equation}
\Xi=(\Xi_1)^{N_1}(\Xi_2)^{N_2}(\Xi_{3a})^{N_{3a}}(\Xi_{3b})^{N_{3b}}...,
\end{equation}
where $N_i=x_iN_s$ is the number of clusters of type $i$. This factorization is
possible because adsorption on a given cluster does not affect any of the
others.  
The adsorption isotherm is, from Eq. \ref{eq:nbar1},  
\begin{equation}
<n> = \lm N_1\left(\frac{\partial\ln \Xi_1}{\partial\lambda}\right)+\lm N_2\left(\frac{\partial\ln \Xi_2}{\partial\lambda}\right)+...
\label{eq:nbarexp}
\end{equation}

\begin{figure}[t]
\begin{center}

\resizebox{8cm}{!}{\includegraphics{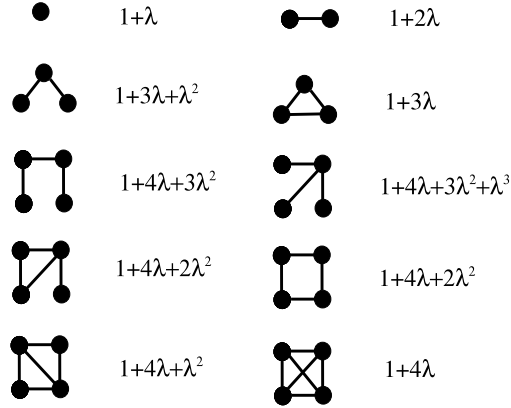}}
\caption{Clusters with grand canonical partition function. A solid line connects two sites that are closer than $\sigma$. 
Triplets of type ``3a'' and ``3b'' are shown left to right in the second row.}\label{fig:clusters}
\end{center}
\end{figure}

Figure \ref{fig:clusters} shows the first few
possible clusters and the associated grand canonical
ensemble partition function. A solid line connecting two sites
indicates that they are closer than a particle diameter and therefore
cannot be simultaneously occupied. The partition function is evaluated
by explicit summation over all sites. As an example, for the first of the two triplets, 
\begin{equation}
\Xi_{3a}=\sum_{n_1=0}^{1}\sum_{n_2=0}^{1}\sum_{n_3=0}^{1}(1-n_1n_2)(1-n_1n_3)\lm^{n_1+n_2+n_3}
=1+3\lm+\lm^2.
\end{equation}

By substituting these results into Eq \ref{eq:nbarexp} we obtain
\begin{equation}
\theta=\alpha\bigl[\frac{\lm}{1+\lm}x_1+\frac{2\lm}{1+2\lm}x_2+\frac{3\lm+2\lm^2}{1+3\lm+\lm^2}x_{3a}+\frac{3\lm}{1+3\lm}x_{3b}+\ldots\bigr]
\end{equation}

\begin{figure}

\begin{center}

\resizebox{8cm}{!}{\includegraphics{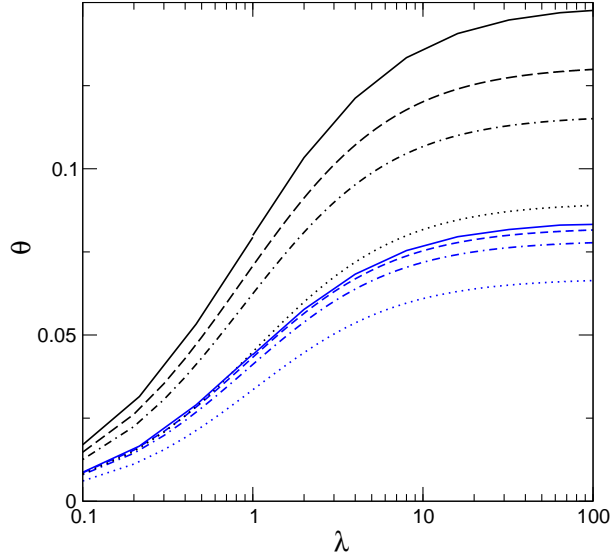}}
\caption{Isotherms for $\alpha=0.2, 0.1$, top to bottom. The solid lines
  show simulation results (an average of 10 distributions of
  4000 sites). The dotted, dash-dotted and dashed lines show the
  cluster expansion to first ($x_1$), second ($x_2$) and third order ($x_{3a}+x_{3b}$), respectively.
}\label{fig:isoc}
\end{center}

\end{figure}

We show the predictions of this equation to different orders compared
with the simulation results in Fig \ref{fig:isoc}. Successive
approximations approach the simulated isotherm from below. Including 
clusters up to triplets described the numerical simulation data accurately for $\alpha=0.1$, but is already quite poor for $\alpha=0.2$. This rapid breakdown can 
be understood by realizing that truncation of the expansion means that
one is failing to include all sites.

An alternative approach is to assume that all sites belong to clusters
up to a given order.  Thus, if all sites are assumed to be isolated we
have

\begin{equation}
\Xi = \Xi_{1}^{N_s}=(1+\lambda)^{N_s},
\end{equation}
which  gives for the coverage the Langmuir expression:
\begin{equation}
\theta_1=\alpha\frac{\lambda}{1+\lambda}.
\end{equation}

\begin{figure}[t]

\begin{center}

\resizebox{8cm}{!}{\includegraphics{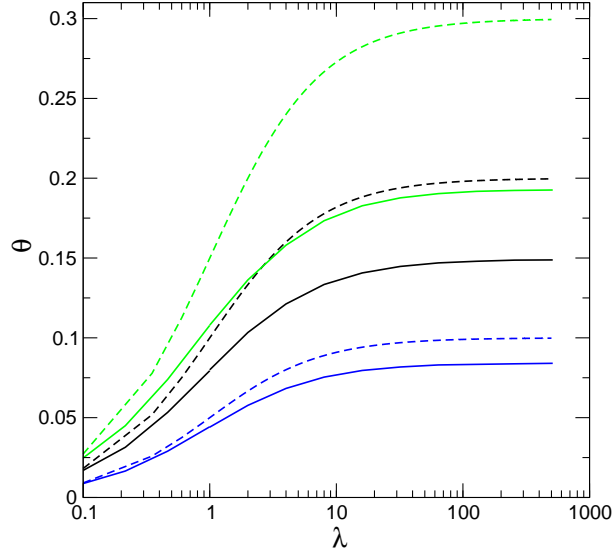}}
\caption{Isotherms for $\alpha=0.3, 0.2, 0.1$, top to bottom. Solid and dashed lines show the simulation
results and  the Langmuir theory, respectively. }\label{fig:iso1}
\end{center}

\end{figure}

For $\lm\to\infty$ this gives $\theta_1=\alpha$, i.e. all sites are occupied. 
When it is assumed that only isolated sites and sites with one
neighbor are present, the number of the latter is estimated as $N_2=(N_s-N_1)/2=N_s(1-\exp(-4\alpha))/2$. 
The partition function is
\begin{equation}
\Xi = \Xi_{1}^{N_1}\Xi_{2}^{N_2}=(1+\lambda)^{N_1}(1+2\lambda)^{N_2},
\end{equation}
which leads to a coverage of
\begin{equation}
\theta_2 =\frac{\alpha}{N_s}\bigr(\frac{N_1\lm}{1+\lm}+\frac{2N_2\lm}{1+2\lm}\bigr)
= \alpha \lambda \frac{1+\lambda(1+\exp(-4\alpha))}{(1+2\lambda)(1+\lambda)}.
\end{equation}

The maximum coverage at this level of approximation, obtained by taking the limit $\lm\to\infty$, is
\begin{equation}
\theta_{2,{\rm max}} =\frac{\alpha}{2}(1+\exp(-4\alpha))
=\alpha -2\alpha^2+O(\alpha^3).
\end{equation}

At the triplet level, we estimate the total number of triplets as
$N_3=(N_s-N_1-2N_2)/3$. Of these,  a fraction $y_{3a}=x_{3a}/(x_{3a}+x_{3b})$ are of type a and the remainder are of type b. This yields:

\begin{equation}
\theta_3 =\alpha\bigr[(y_{3b}\frac{3\lm}{1+3\lm}+y_{3a}\frac{3\lm+2\lm^2}{1+3\lm+\lm^2})\frac{1-x_1-2x_2}{3}+\frac{2\lm}{1+2\lm}x_2+\frac{\lm}{1+\lm}x_1\bigr].
\end{equation}

The maximum coverage is then:
\begin{equation}
\theta_{3,{\rm max}} =\alpha[(y_{3a}+1)(1-x_1-2x_2)/3+x_2+x_1].
\end{equation}

\begin{figure}[t]
\begin{center}

\resizebox{8cm}{!}{\includegraphics[bb=29 48 545 522]{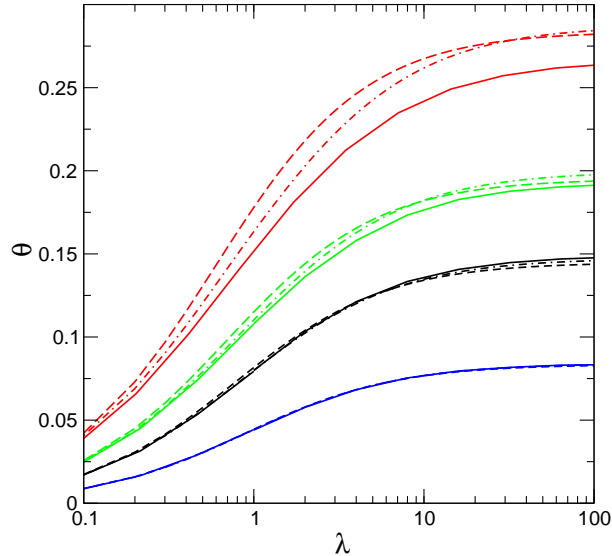}}
\caption{Isotherms for $\alpha=0.5, 0.3, 0.2, 0.1$,  top-to-bottom. The solid lines show the simulation results, 
while the dashed and dash-dot lines show the approximate theories 
$\theta_2$ and $\theta_3$, respectively}\label{fig:iso2}
\end{center}
\end{figure}

Figures \ref{fig:iso1} and \ref{fig:iso2} show the simulated isotherms
compared with the alternative theory. Unlike the theoretical estimates shown in
Fig \ref{fig:isoc}, the estimates generally lie above the simulation results. 
It is clear that the Langmuir model (all sites
assumed to be isolated) provides a very poor description, even at
quite low site densities. There is not much difference between the
second and third order cluster expansions. Both provide a satisfactory
description of the isotherm for $\alpha\leq 0.3$.

\section{Conclusion}

We have obtained numerical results for the isotherms of hard spheres
adsorbing on a random site surface. For a given configuration of sites
the maximum coverage is obtained by taking the limit of infinite bulk
phase activity. We developed an approximate theory based on a cluster
expansion of the partition function that is accurate up to moderate site
density. 

Weight and Hartmann obtained an analytic solution for a related model
of hard spheres on a random graph using replica theory.  Their
solution applies only to the densest particle packing.  Generalization
of their approach to finite activity or to the RSS surface does not
appear to be straightforward. Perhaps an alternative approach will
allow a description of the thermodynamics for a wider range of site
densities than is possible using the theory proposed here.

\section*{Acknowledgments}
We thank David Dean, Gilles Tarjus and Pascal Viot for useful discussions. 

\end{document}